\documentclass[prl,aps,twocolumn,twoside,superscriptaddress]{revtex4}
\usepackage{amssymb}
\usepackage{amsmath}
\usepackage{amscd}
\usepackage{latexsym}
\usepackage{epsfig}
\usepackage{graphicx}
\usepackage{bbm}

\def\QED{\mbox{\rule[0pt]{1.5ex}{1.5ex}}}

\def\endproof{\hspace*{\fill}~\QED\par\endtrivlist\unskip}

\def\Tr{\mathop{\rm Tr}\nolimits}

\newcommand{\defeq}{\stackrel{\rm def}{=}}

\def\CC{\mathop{\rm CC}}
\newcommand{\cH}{{\cal H}}

\newcommand{\cK}{{\cal K}}
\newcommand{\cS}{{\cal S}}

\newcommand{\cX}{{\cal X}}

\newcommand{\rhon}{\rho^{\otimes n}}

\def\bx{{\bm x}}

\def\Op{\kappa}
\begin{document}
\title{Optimal Visible Compression Rate For Mixed States 
\\
Is Determined By Entanglement Of Purification}
\author{Masahito Hayashi}
\email{masahito@qci.jst.go.jp}
\affiliation{ERATO-SORST 
Quantum Computation and Information Project, JST, Tokyo 113-0033, Japan}
\affiliation{Superrobust Computation Project,
Information Science and Technology Strategic Core (21st Century COE by MEXT), 
Graduate School of Information Science and Technology,
The University of Tokyo }
\begin{abstract}
Entanglement of purification was introduced by 
Terhal {\it et al.}\cite{THLD} for characterizing 
the bound of the generation of correlated states from
maximally entangled states 
with sublinear size of classical communication.
On the other hand, 
M. Horodecki \cite{M-Ho2} obtained 
the optimal compression rate
with a mixed states ensemble in the visible setting.
In this paper, we prove that
the optimal visible compression rate for mixed states
is equal to the limit of the regularized entanglement of 
purification of the state corresponding to the given ensemble.
This result gives a new interpretation to
the entanglement of purification.
\end{abstract}

\maketitle
{\it Introduction:~}
Many information quantities appear 
as the bound of the respective information processing
in quantum information theory.
Usually, these bounds depend on the information processing
of interest.
However, recently,
Devetak {\it et al.} \cite{DHW}
considered the relation among quantum information processing.
In their paper, they found remarkable conversions among
quantum protocols,
and succeeded in simplifying the proofs
of several important theorems.
Also Bennett {\it et al.} \cite{BDSW}
obtained the conversion relation among
error correction of quantum channel 
and one-way distillation of mixed entangled state.
In this paper, 
we focus on
visible quantum data compression with mixed states and 
generation of a correlated state from maximally entangled state
with classical communication of the sublinear size.
Using a conversion relation similar to Bennett {\it et al.} \cite{BDSW},
we find
an interesting conversion relation between 
the quantum compression with a mixed states ensemble and 
the state generation from maximally entangled state.

Quantum data compression 
was initiated by Schumacher\cite{Schumacher}.
As the quantum information source,
he focused on the 
quantum states ensemble $(p_x,W_x)_{x \in \cX}$,
in which the quantum state $W_x$ generates with the probability $p_x$.
He showed that 
the asymptotic optimal compression rate $R(W,p)$
is equal to the entropy $H(W_p)\defeq - \Tr W_p \log W_p$ of
the average state $W_p\defeq \sum_x p_x W_x$
of this ensemble.
In his original problem,
the encoder is restricted to performing a quantum operation.
However, M. Horodecki \cite{M-Ho1} considered
another problem, 
in which the encoder is defined as the any map from 
$\cX$ to the quantum states.
This formulation is called visible,
while the former is called blind.
He also showed that even in the visible setting
if every state $W_x$ is pure,
the optimal rate $R(W,p)$ is equal to the entropy rate 
$H(W_p)$.
However, it had been an open problem
to characterize the rate $R(W,p)$ in the mixed states case.
M. Horodecki\cite{M-Ho2} studied this problem,
and succeeded in its characterization.
However, his characterization contains a limiting expression.
Hence, it is an open problem whether
it can be characterized without any limiting expression.

On the other hand,
Terhal {\it et al.}\cite{THLD}
introduced entanglement of purification $E_p(\rho)$ for 
any partially entangled state $\rho$.
They also consider the 
generation of the tensor product of any partially entangled state 
$\rho$ on the composite system $\cH_A \otimes \cH_B$
from 
maximal entangled states in the asymptotic form.
In particular, they restrict the rate 
of the classical communication to zero asymptotically.
Indeed, when the target state $\rho$ is pure,
this optimal rate is $H(\Tr_A\rho)$,
which is equal to the optimal rate without any restriction for 
the rate of the classical communication\cite{LP2}.
Their main result is that 
the optimal rate with this restriction 
is equal to 
$\lim_{n \to \infty} \frac{E_p(\rho^{\otimes n})}{n}$.
Of course, if the entanglement of purification satisfies 
the additivity, {\it i.e.}, $E_p(\rho)+E_p(\sigma)=E_p(\rho\otimes \sigma)$,
this optimal rate is equal to the entanglement of purification.
However, this additivity is still open.

In this paper, 
we give another formula for 
the optimal visible compression rate
$R(W,p)$ as
\begin{align}
R(W,p)
= \lim_{n \to \infty} \frac{1}{n} E_p(\tilde{W}_p^{\otimes n}),\label{2-9-1} 
\end{align}
where $\tilde{W}_p 
\defeq \sum_x p_x |e_x^A\rangle \langle e_x^A|\otimes W_x$.
In order to prove this equation,
we first give an error-free visible compression protocol of
the specific ensemble related to 
the state generated by LOCC from a maximally entangled state,
which is close to $\tilde{W}_p^{\otimes n}$.
This compression is realized by quantum memory 
the size of a maximally entangled state and 
classical communication with the same size as the LOCC operation.
Next, we prove this ensemble is close to our target ensemble.
Combining them, we prove that the optimal compression rate is 
less than the regularized entanglement of purification
$\lim_{n \to \infty} \frac{1}{n} E_p(\tilde{W}_p^{\otimes n})$
The converse relation is proved from the axiomatic properties of 
entanglement of purification.
Hence, using the relation (\ref{2-9-1}), 
we clarify the relation between 
the two problems,
the mixed state compression and 
the state generation from maximally entangled state 
with classical communication of the sublinear size.
Thus, if the additivity of entanglement of purification is proved,
the optimal rate of visible compression is equal to 
entanglement of purification
of the state corresponding to the given ensemble.

\medskip
{\it State generation from maximally 
entangled state with communication of the sublinear size:~}
In state generation from maximally entangled state,
our protocol is described by 
an LOCC quantum operation (TP-CP) $\kappa$ and
the initial maximally entangled state 
$|\Phi_{L}\rangle\langle \Phi_{L} |$ with the size $L$.
When we generate 
a partially entangled state $\rho$ on the composite system
$\cH_A \otimes \cH_B$ by this protocol,
its performance is evaluated by 
i) the size $L$,
ii) the quality of the generated state,
which is given by 
\begin{align*}
\varepsilon(\rho,\kappa,L)\defeq
1- F^2(\rho,\Op(|\Phi_{L}\rangle\langle \Phi_{L} |) ),
\end{align*}
where $F(\rho,\sigma)$ is the fidelity $\Tr | \sqrt{\rho}\sqrt{\sigma}|$,
and iii) the size of classical communication,
which is denoted by $\CC(\kappa)$.

In the asymptotic formulation, 
the bound with classical communication of the sublinear size
is given by
\begin{align*}
E_c^{\dashrightarrow}(\rho)
\defeq 
\inf_{ \{\Op_n,L_n\}} \left\{\left.
\varlimsup   \frac{\log L_n}{n}
\right|
\begin{array}{l}
\varepsilon(\rho^{\otimes n},\kappa_n,L_n)\to 0\\
\frac{\log \CC(\kappa_n)}{n}\to 0
\end{array}
\right\} .
\end{align*}
The RHS is the infimum value of 
$\varlimsup   \frac{\log L_n}{n}$
under the conditions
$\varepsilon(\rho^{\otimes n},\kappa_n,L_n)\to 0$ and 
$\frac{\log \CC(\kappa_n)}{n}\to 0$.
Lo \& Popescu \cite{LP2} 
calculated this value in the pure states case as follows.
\begin{align*}
E_c^{\dashrightarrow}(|u \rangle \langle u|)
= H(\Tr_B |u \rangle \langle u|).
\end{align*}
Further, Terhal {\it et al}. \cite{THLD}
introduced the entanglement of purification $E_p(\rho)$ 
as
\begin{align*}
E_p(\rho)\defeq \min_{u:\Tr_{A_2,B_2}| u\rangle \langle u|= \rho}
H(\Tr_B |u \rangle \langle u|),
\end{align*}
where $\cH_{A_2}$ and $\cH_{B_2}$ are additional spaces.
In the above definition, 
$u$ is a purification of $\rho$ with the reference system
$\cH_{A_2}\otimes \cH_{B_2}$.
$\Tr_B$ is the partial trace concerning the original space $\cH_B$ and
the additional space $\cH_{B_2}$.

Using Lo \& Popescu's result,
they showed that
\begin{align}
E_c^{\dashrightarrow}(\rho)
\ge E_p(\rho).
\end{align}
Applying this inequality to $\rhon$,
they also showed
\begin{align}
E_c^{\dashrightarrow}(\rho)
\ge \lim_{n \to \infty} \frac{E_p(\rhon)}{n}.
\end{align}
Further, they proved the following properties
for entanglement of purification:
\begin{description}
\item[\bf E1](Normalization) 
$E_p(\rho)=\log d$ when $\rho$ is a maximally entangled state 
of dimension $d$.
\item[\bf E2](Weak monotonicity)
Let $\kappa$ be a operation containing quantum communication with size
$d$. Then,
\begin{align*}
E_p(\kappa(\rho))
\le E_p(\rho)+ \log d.
\end{align*}
\item[\bf E3](Continuity)
When any sequences of two states 
\{$\rho_n\}$ and $\{\sigma_n\}$ on the system $\cH_n$
satisfy $\|\rho_n - \sigma_n\|_1\to 0$, the convergence
$\frac{|E_p(\rho_n) - E_p(\sigma_n)|}{\log \dim \cH_n}\to 0$ holds.
\item[\bf E4](Convergence)
The quantity $\frac{E_p(\rhon)}{n}$ converges as $n \to \infty$.
\end{description}

Using these properties, they showed the opposite inequality
\begin{align}
E_c^{\dashrightarrow}(\rho)
\le \lim_{n \to \infty} \frac{E_p(\rhon)}{n}.
\end{align}
Hence, we obtain the relation
\begin{align}
E_c^{\dashrightarrow}(\rho)
= \lim_{n \to \infty} \frac{E_p(\rhon)}{n}\label{th-1}.
\end{align}

Further, they obtained the following property:
\begin{align}
E_p(\rho)\le H(\rho^A).\label{th-2}
\end{align}

\medskip
{\it Visible State Compression:~}
In the visible state compression,
we consider the compressed quantum system $ \cK $.
The encoder is given by a map $\tau$ from 
$\cX$ to $\cS(\cK)$,
and 
the decoder is represented by a TP-CP map 
$\nu$ from $\cS(\cK)$ to $\cS(\cH)$.  
The triple $\Psi \defeq (\cK,\tau,\nu)$ 
is called a visible code. 
That is, the information is stored by a quantum memory.
Therefore, 
the error
$\varepsilon_p(\Psi)$
and the size $|\Psi|$ of the code $\Psi$ 
are defined as follows:
\begin{align*}
\varepsilon_p(\Psi) & \defeq \sum_{x \in \cX}p_x 
\left(1- F^2(W_x,\nu \circ \tau (x)) \right) ,\quad
|\Psi|\defeq \dim \cK .
\end{align*}
Then, the optimal compression rate is given by
\begin{align*}
R(W,p)& \defeq 
\inf_{\{\Psi^{(n)}\}}
\left\{\left. \varlimsup \frac{1}{n} \log | \Psi^{(n)}|
\right|
\varepsilon_{p^n} (\Psi^{(n)}) \to 0
\right\}.
\end{align*}
Indeed, if the encoder $\tau$ is given as a TP-CP map (quantum operation),
the setting is called blind.
In order to treat this problem,
M. Horodecki \cite{M-Ho2} focused on the quantity:
\begin{align*}
H^{ext}(W,p)\defeq
\inf_{ W^{ext}_x: {\rm purification~of~} W_x} 
H(\sum_x p_x W^{ext}_x),
\end{align*}
and showed 
\begin{align}
R(W,p)= \lim_{n \to \infty} \frac{H^{ext}(W^{(n)},p^n)}{n}. \label{9-17-28}
\end{align}
The following is the main theorem.

{\it Theorem:}
The optimal compression rate is given by 
\begin{align}
R(W,p)
&= E_c^{\dashrightarrow}(\tilde{W}_p)
= \lim_{n \to \infty} \frac{1}{n} E_p(\tilde{W}_p^{\otimes n}), 
\label{9-17-29}\\
\tilde{W}_p&\defeq \sum_x p_x |e_x^A\rangle \langle e_x^A|\otimes W_x,
\nonumber
\end{align}
where the $\{e_x^A\}$ is CONS indicated by $x \in \cX$.

From the definition of $E_p(\tilde{W}_p)$,
we can easily check that $E_p(\tilde{W}_p)\le H^{ext}(W,p)$.
Using this theorem, we obtain 
\begin{align}
\lim_{n \to \infty} \frac{H^{ext}(W^{(n)},p^n)}{n}
=\lim_{n \to \infty} \frac{E_p(\tilde{W}_p^{\otimes n})}{n} .
\end{align}
Further, when all states $W_x$ are pure, 
we obtain $\lim_{n \to \infty} \frac{1}{n} E_p(\tilde{W}_p^{\otimes n})
=H(W_p)$,
which implies 
$H(W_p)\le E_p(\tilde{W}_p)$.
From (\ref{th-2}), we have
\begin{align}
E_p(\tilde{W}_p)= H(W_p).
\end{align}

\medskip
{\it Proof of direct part:\quad}
In this paper, the direct part
means the existence of the visible compression attaining
the limit of the regularized entanglement of purification
of the state corresponding to the given ensemble
while
the converse part does 
the nonexistence of 
the visible compression 
with a smaller rate than 
the limit of the regularized entanglement of purification
of the state corresponding to the given ensemble.
The direct part follows the following lemma.

We briefly mention our construction of a code $\Psi$
before going to its detail.
First, we choose a one-way LOCC operation $\kappa$
such that the state 
$\kappa(|\Phi_L \rangle\langle \Phi_L|)$
is close to $\tilde{W}_p$.
Assume that we perform the measurement 
$\{|e_x^A\rangle \langle e_x^A|\otimes I\}_x$.
When the state is $\tilde{W}_p$,
the final state on $\cH_B$ with the measurement outcome 
$x$ is $W_x$.
Hence, when the state is $\kappa(|\Phi_L \rangle\langle \Phi_L|)$,
we can expect that 
the final state $W_x'$
on $\cH_B$ with the outcome 
$x$ is close to $W_x$.
Further, the ensemble $(W_x')_{x \in \cX}$
can be compressed 
to the pair of 
classical information of the size $\CC(\kappa)$
and 
Hilbert space of the dimension $L$ in the visible framework
without any error.
When this compression protocol is described by a code $\Psi$,
this insight is formulated as the following lemma.

{\it Lemma:}
Let $\kappa$ be a one-way LOCC operation.
There exists a code $\Psi$ such that
\begin{align}
\frac{1}{2} \varepsilon_p(\Psi)
&\le 
1-F^2(\tilde{W}_p, \kappa(|\Phi_L\rangle \langle\Phi_L |))
\nonumber \\&\quad 
+\frac{1}{2}
\|\tilde{W}_p- \kappa (|\Phi_L\rangle \langle \Phi_L|)\|_1
,\label{11-21-2-q}\\
|\Psi|&= L \cdot \CC(\kappa). \label{11-21-6-q}
\end{align}
(Note that any two-way LOCC operation can be simulated by one-way LOCC
when the initial state is pure \cite{LP}.)

Using this lemma, we obtain the 
direct part as follows.
Let $\kappa_n$ be a one-way LOCC operation 
satisfying 
\begin{align*}
&\lim_{n \to \infty} F(\tilde{W}_p^{\otimes n}, \kappa_n 
(|\Phi_{L_n}\rangle \langle \Phi_{L_n}|))=1,~
\frac{\log \CC(\kappa_n)}{n} \to 0,\\
&\lim_{n \to \infty} \frac{\log L_n}{n} 
\le E_c^{\dashrightarrow}(\tilde{W}_p)+\epsilon
\end{align*}
for any $\epsilon > 0$.
Thus, by applying this lemma, there exists 
a sequence of codes $\{\Psi_n\}$ such that
$\varepsilon_{p^n}(\Psi_n)\to 0$ and 
$\lim_{n \to \infty} \frac{\log |\Psi_n|}{n} \le 
E_c^{\dashrightarrow}(\tilde{W}_p)+\epsilon$.
Therefore, we obtain 
$R(W,p)\le E_c^{\dashrightarrow}(\tilde{W}_p)$.

\medskip
{\it Construction of the code $\Psi$ satisfying (\ref{11-21-2-q})
and (\ref{11-21-6-q}):~}
The following construction of $\Psi$ from one-way LOCC operation $\kappa$
is similar to 
a simulation of 
one-way LOCC distillation protocol 
by quantum error correction\cite{BDSW}.
We give an error-free visible compression protocol of
the ensemble 
$
\left(\frac{\Tr_B (|e_x^A\rangle \langle e_x^A|\otimes I_B)
\kappa(|\Phi_L \rangle\langle \Phi_L|)}
{\Tr (|e_x^A\rangle \langle e_x^A|\otimes I_B)
\kappa(|\Phi_L \rangle\langle \Phi_L|)}
\right)_{x\in \cX}$
with the compression size
$L \cdot \CC(\kappa)$.
Assume that the operation $\kappa$ has the form
$\kappa= \sum_i \kappa_{A,i} \otimes \kappa_{B,i}$,
where $\{ \kappa_{A,i} \}_{i=1}^{l_n}$ is an instrument 
(CP maps valued measure) on $\cH_A$ 
and $\kappa_{B,i}$ is a TP-CP map on $\cH_B$ for each $i$.
Define the probability $q_x$ 
\begin{align}
q_x \defeq & 
\Tr (|e_x^A\rangle \langle e_x^A|\otimes I_B)
\Bigl(\sum_i \kappa_{A,i} \otimes \kappa_{B,i}
(|\Phi_L \rangle\langle \Phi_L|)\Bigr)
\label{11-21-8-q}\\
=&\sum_i \Tr \kappa_{A,i}^* 
((|e_x^A\rangle \langle e_x^A|)\otimes I_B
(|\Phi_L \rangle\langle \Phi_L|))
\nonumber 
\end{align}
and 
the probability $p_{i,x}$ and the state $\rho_{i,x}$ as
\begin{align*}
p_{i,x} & \defeq 
\frac{\Tr \kappa_{A,i}^* ((|e_x^A\rangle \langle e_x^A|)\otimes I_B
(|\Phi_L \rangle\langle \Phi_L|))}{q_x}\\
\rho_{i,x} & \defeq
\frac{\Tr_A 
\kappa_{A,i}^* ((|e_x^A\rangle \langle e_x^A|)\otimes I_B
(|\Phi_L \rangle\langle \Phi_L|))}{q_x p_{i,x}}.
\end{align*}
Now, we construct the coding protocol $\Psi$:
When the encoder receives the input signal $x$,
he sends the state $\rho_{i,x}$ with the probability $p_{i,x}$
and sends the classical information $i$.
The decoder performs the TP-CP map $\kappa_{B,i}$ dependently of 
the classical signal $i$.
This protocol gives the visible compression 
the ensemble 
$
\left(\frac{\Tr_B (|e_x^A\rangle \langle e_x^A|\otimes I_B)
\kappa(|\Phi_L \rangle\langle \Phi_L|)}
{\Tr (|e_x^A\rangle \langle e_x^A|\otimes I_B)
\kappa(|\Phi_L \rangle\langle \Phi_L|)}
\right)_{x\in \cX}$,
which can be realized by 
classical memory of the size $\CC(\kappa)$
and 
quantum memory with the dimension $L$.
That is, inequality (\ref{11-21-6-q}) follows from this construction.

Next, we prove that the ensemble 
$
\left(p_x,\frac{\Tr_B (|e_x^A\rangle \langle e_x^A|\otimes I_B)
\kappa(|\Phi_L \rangle\langle \Phi_L|)}
{\Tr (|e_x^A\rangle \langle e_x^A|\otimes I_B)
\kappa(|\Phi_L \rangle\langle \Phi_L|)}
\right)_{x\in \cX}$
is close to the given ensemble,
{\it i.e.}, show inequality (\ref{11-21-2-q}).
This inequality follows from the evaluation:
\begin{align}
& F^2
\Bigl(\sum_{x} p_{x}|e_x^A\rangle \langle e_x^A|\otimes W_x,
\sum_i \kappa_{A,i} \otimes \kappa_{B,i}
(|\Phi_L \rangle\langle \Phi_L|)\Bigr)\nonumber \\
\le &
\Tr \sqrt{\sum_{x} p_{x}|e_x^A\rangle \langle e_x^A|\otimes W_x}
\sqrt{\sum_i \kappa_{A,i} \otimes \kappa_{B,i}
(|\Phi_L \rangle\langle \Phi_L|)}\nonumber \\
= &
\Tr \sum_{x} \sqrt{p_{x}}|e_x^A\rangle \langle e_x^A|\otimes \sqrt{W_x}
\sqrt{
\sum_i \kappa_{A,i} \otimes \kappa_{B,i}
(|\Phi_L \rangle\langle \Phi_L|)}\nonumber \\
= &
\sum_{x} \sqrt{p_{x}}
\Tr_B \Bigl[\sqrt{W_x}\nonumber \\
& \cdot
\Tr_A \Bigl(|e_x^A\rangle \langle e_x^A|\otimes I_B\Bigr)
\sqrt{
\sum_i \kappa_{A,i} \otimes \kappa_{B,i}
(|\Phi_L \rangle\langle \Phi_L|)}
\Bigl]\nonumber \\
\le &
\sum_{x} \sqrt{p_{x}}
\Tr_B \Bigl[\sqrt{W_x} \nonumber \\
&\cdot \sqrt{
\Tr_A \Bigl(|e_x^A\rangle \langle e_x^A|\otimes I_B \Bigr)
\sum_i \kappa_{A,i} \otimes \kappa_{B,i}
(|\Phi_L \rangle\langle \Phi_L|)\Bigr)}\Bigr]\nonumber \\
= &
\sum_{x} \sqrt{p_{x} q_x}
\Tr_B \sqrt{W_x} 
\sqrt{\sum_i p_{i,x} \kappa_{B,i} (\rho_{i,x})}\label{9-8-1}\\
= &
\sum_{x} (\sqrt{p_{x} q_x} - p_{x})
\Tr_B \sqrt{W_x} 
\sqrt{\sum_i p_{i,x} \kappa_{B,i} (\rho_{i,x})}
\nonumber \\
&+
\sum_{x} p_{x}
\Tr_B \sqrt{W_x} 
\sqrt{\sum_i p_{i,x} \kappa_{B,i} (\rho_{i,x})}.
\label{991}
\end{align}
The above relations can be checked as follows:
i) The first inequality follows
from a basic inequality $F^2(\rho,\sigma)\le\Tr \sqrt{\rho}\sqrt{\sigma}$.
ii) The second inequality follows from 
the matrix concavity of $\sqrt{t}$.
iii) The equation (\ref{9-8-1}) follows from 
\begin{align*}
& q_x \sum_i p_{i,x} \kappa_{B,i} (\rho_{i,x})\\
= &
\sum_i 
\kappa_{B,i}
(\Tr_A 
(\kappa_{A,i}^* (|e_x^A\rangle \langle e_x^A|)\otimes I_B)
|\Phi_L \rangle\langle \Phi_L|)\\
= &
\Tr_A |e_x^A\rangle \langle e_x^A|\otimes I_B
\sum_i \kappa_{A,i} \otimes \kappa_{B,i}
(|\Phi_L \rangle\langle \Phi_L|),
\end{align*}
where $\iota_B$ is the identical operation on $\cH_B$.
The second term of (\ref{991}) is evaluated by
\begin{align}
& \sum_{x} (\sqrt{p_{x} q_x} - p_{x})
\Tr_B \sqrt{W_x} 
\sqrt{\sum_i p_{i,x} \kappa_{B,i} (\rho_{i,x})} \nonumber \\
\le &
\sum_{x} (\sqrt{p_{x} q_x}-p_x)_+ 
=  \sum_{x} \bigl(\sqrt{\frac{q_{x}}{p_x}}-1\bigr)_+ p_x \nonumber \\
\le &\sum_{x} (\frac{q_{x}}{p_x}-1)_+ p_x 
= \sum_x (q_x - p_x)_+
= \frac{1}{2}\|q -p \|_1 \nonumber \\
\le &
\frac{1}{2}
\|\tilde{W}_p- \kappa (|\Phi_L\rangle \langle \Phi_L|)\|_1,
\label{11-21-3-q}
\end{align}
where
$(t)_+$ is $t$ when $t$ is positive and it is $0$ otherwise.
The final inequality follows from the definition of the distribution $q$
(\ref{11-21-8-q}).

Concerning the first term of (\ref{991}), 
the inequality
\begin{align}
& \frac{1}{2}(1- F^2(W_x,\sum_i p_{i,x} \kappa_{B,i} (\rho_{i,x})))
\nonumber \\
\le & 1- F(W_x,\sum_i p_{i,x} \kappa_{B,i} (\rho_{i,x}))\nonumber\\
\le & 1- \Tr_B \sqrt{W_x} 
\sqrt{\sum_i p_{i,x} \kappa_{B,i} (\rho_{i,x})}
\label{11-21-4-q}
\end{align}
holds.
Hence, 
(\ref{11-21-2-q}) 
follows from (\ref{991}), (\ref{11-21-3-q}), and (\ref{11-21-4-q}).

\endproof

\medskip
{\it Proof of converse part:~}
The converse part essentially follows from Conditions {\bf E2} and {\bf E3}
of entanglement of purification.
For any $\epsilon >0$, we choose a sequence of codes 
$\Psi_n=(\cK_n,\tau_n,\nu_n)$
such that
\begin{align*}
R \defeq \varlimsup \frac{1}{n}\log |\Psi_n|
\le R(W,p)+ \epsilon, \quad
\varepsilon_{p^n}(\Psi_n) \to 0.
\end{align*}
The state $\tilde{W}_p^{\otimes n}$ is described by 
the i.i.d. distribution $\{p_x^n\}_{x \in \cX^n}$ 
of $\{p_x\}_{x \in \cX}$
as $\tilde{W}_p^{\otimes n}
= \sum_{{\bx} \in \cX^n} p^n_{\bx}|e^A_{\bx}\rangle \langle e^A_{\bx}|
\otimes W^{(n)}_{\bx}$.
Then, the state $\tilde{\rho}_n\defeq
\sum_{{\bx} \in \cX^n}p^n_{\bx}|e^A_{\bx}\rangle \langle e^A_{\bx}|
\otimes \tau_n(\bx)$
satisfies 
\begin{align*}
& F(\tilde{W}_p^{\otimes n},\iota_A\otimes \nu_n (\tilde{\rho}_n))
=
\sum_{{\bx} \in \cX^n}p^n_{\bx}
F(W^{(n)}_{\bx},\nu_n \circ \tau_n(\bx)) \\
\ge &
\sum_{{\bx} \in \cX^n}p^n_{\bx}
F^2(W^{(n)}_{\bx},\nu_n \circ \tau_n(\bx))\to 1,\label{9-10-6}
\end{align*}
where $\iota_A$ is the identical operation on the system $\cH_A$.
Note that $\tau_n$ is the encoding and $\nu_n$ is the decoding.
From (\ref{th-2}) and 
Condition {\bf E2}(weak monotonicity of entanglement of purification),
\begin{align*}
\log |\Psi_n| \ge H(\Tr_A \tilde{\rho}_n)
\ge E_p(\tilde{\rho}_n)
\ge E_p(\iota_A\otimes \nu_n (\tilde{\rho}_n)).
\end{align*}
Hence, Condition {\bf E3}(continuity of entanglement of purification)
 yields
\begin{align*}
\varliminf \frac{1}{n} \log |\Psi_n|
\ge \lim_{n \to \infty} \frac{1}{n} E_p(\tilde{W}_p^{\otimes n}).
\end{align*}
Hence, using (\ref{th-1}), we obtain
\begin{align*}
R(W,p)\ge E_c^{\dashrightarrow}(\tilde{W}_p)
= \lim_{n \to \infty} \frac{1}{n} E_p(\tilde{W}_p^{\otimes n}).
\end{align*}
\endproof

\medskip
{\it Conclusion:~}
We have proved that 
the bound of visible mixed state compression is equal to 
the optimal bound of 
the state generation from maximally 
entangled state with classical communication of the sublinear size.
In particular, inequalities (\ref{11-21-2-q}) and (\ref{11-21-6-q})
express the relation between two problems.
In the proof of direct part,
we have constructed an error-free visible compression protocol
of the ensemble corresponding to 
a state generation protocol from maximally entangled state.
The converse part has been proved from the weak monotonicity and 
continuity of entanglement of purification.
This result may indicate
that these two problems are essentially equivalent.
The obtained relation is essentially based on the relation
between the noiseless channel and the maximally entangled state.
Hence, a further relation based on this relation 
can be expected among 
several information protocols.

Further, when there is no restriction 
concerning the size of classical communication,
the optimal rate of the state generation from maximally entangled state 
(entanglement cost)
is closely related to additivity of the channel capacity\cite{MSW,Shor}.
Hence, it is interesting 
to consider the relation between
state generation with sublinear-size classical communication
and channel problems.

The author thanks
Professor Hiroshi Imai and 
ERATO-SORST 
Quantum Computation and Information Project
for supporting this research.
He is also grateful for Dr. Andreas Winter to useful discussion on 
the related topics.
He is also indebted to the reviewer for commenting this paper
and pointing out the relation with the paper \cite{BDSW}.
 	Help
Submitted manuscripts - follow the links to view or make changes and resubmit

LL10788 	View	Optimal Visible Compression Rate For Mixed States Is De

\end{document}